\documentclass[12pt,a4paper]{article}
\usepackage{graphics}

\newcommand\as{\alpha_{\scriptscriptstyle{S}}}

\begin{document}
\begin{titlepage}
\begin{flushright}
  RAL-TR-1999-051 \\
  BICOCCA-FT-99-24 \\
  hep-ph/9908388
\end{flushright}
\par \vspace{10mm}
\begin{center}
{\Large \bf
Initial State Radiation in Simulations\\[1ex]
of Vector Boson Production\\[2ex]
at Hadron Colliders}
\end{center}
\par \vspace{2mm}
\begin{center}
{\bf G. Corcella $^{1,2}$ and M.H. Seymour $^2$}\\
\vspace{2mm}
{$^1$ Dipartimento di Fisica, Universit\`a di Milano
  and INFN, Sezione di Milano}\\
{Via Celoria 16, I-20133, Milano, Italy}
\par \vspace{2mm}
\vspace{2mm}
{$^2$ Rutherford Appleton Laboratory, Chilton,}\\
{Didcot, Oxfordshire.  OX11 0QX\@.  U.K.}
\end{center}

\par \vspace{2mm}
\begin{center} {\large \bf Abstract} \end{center}
\begin{quote}
\pretolerance 10000
The production of vector bosons at present and future hadron colliders
will provide a crucial test for QCD and Standard Model physics.
In this paper we improve parton shower simulations of Drell--Yan processes
by implementing matrix-element corrections to the initial-state radiation.
We apply our work to the HERWIG Monte Carlo event generator and compare
our phenomenological results with the ones obtained using the previous
version of HERWIG, with resummed calculations which we match to the exact
first-order perturbative result, and with recent
Tevatron data. We also make some predictions for jet events at the LHC.
\end{quote}

\vspace*{\fill}
\begin{flushleft}
  RAL-TR-1999-051 \\
  BICOCCA-FT-99-24 \\
  August 1999
\end{flushleft}
\end{titlepage}

\section{Introduction}

The production of vector bosons $W^{\pm}$, $Z^0$ and $\gamma^*$
[\ref{altarelli}] in high energy hadronic collisions is one
of the most important processes that should be
investigated in order to test the Standard Model of electroweak interactions
and Quantum Chromodynamics. By measuring the rapidity or
the mass distribution of the leptonic decay products one can also investigate
the quark and antiquark distributions inside the colliding hadrons.
For such processes, generated at the lowest
order via a hard scattering like $q\bar q'\to V$, where $q$ and $q'$ have the
same flavour for $Z/\gamma^*$ production and different flavour for
$W$ production, higher order corrections
due to multiple gluon emission in the initial-state radiation will play a
crucial role. Many analyses have been devoted to the phenomenology of
vector boson production, particularly to the differential
distribution with respect to the transverse momentum $q_T$ of the produced
vector boson.
The approach of resummation of large logarithms of the ratio $m_V/q_T$
has been followed in many cases.
This was originally proposed by Dokshitzer, Dyakonov and Troyan
(DDT) [\ref{ddt}] and then accomplished by Collins, Soper and Sterman
(CSS) [\ref{collins}], who performed the leading logarithmic
resummation in the space of the impact parameter $b$, which is the
Fourier space conjugate to~$q_T$.
Ladinsky and Yuan implemented the CSS results numerically [\ref{ly}].
Resummations of the initial-state multiple emission 
have been performed in [\ref{davies}] in the $b$-space and more recently in
[\ref{arnold}--\ref{stirling}]
in both the $b$- and the $q_T$-space.
In [\ref{arnold},\ref{ellis1},\ref{ellis2}] the resummation is also matched 
with the exact perturbative first-order result, which is
important at high $q_T$.
In the $b$-space approach non-perturbative effects in the region of large
values of $b$ are taken into account via Gaussian functions in $b$,
corresponding to a smearing of the transverse momentum distribution
[\ref{collins},\ref{ly}], which can also be directly implemented in
$q_T$-space [\ref{ellis2}].

Another possible approach to studying the phenomenology of vector bosons
is to use Monte Carlo simulations of the initial-state parton shower.
Standard parton showers [\ref{pythia},\ref{herwig}] are performed in the
leading-log approximation, therefore they are reliable only in the soft or
collinear region of the phase space, corresponding to low $q_T$ values for
the produced vector boson.
If we wish to study the high $q_T$ region of the spectrum it is necessary to
provide parton showers with matrix-element corrections.
Refs.~[\ref{miu},\ref{mrenna}] implement matrix-element corrections to
simulations of
vector boson production in the PYTHIA Monte Carlo event generator and compare
them with the results obtained at the Tevatron collider by
the D\O\ collaboration.

In this paper we reconsider this problem and apply matrix-element corrections
to the initial-state radiation of the HERWIG parton shower, following the
general prescription contained in [\ref{sey1}], as we already did for
$e^+e^-$ annihilation [\ref{sey2}], Deep Inelastic Scattering [\ref{sey3}]
and top quark decays [\ref{corcella}].

It is worth recalling that at present no Monte Carlo program including
the full next-to-leading order (NLO)
results exists, as it is not known how to set up a
full NLO calculation in a probabilistic way. When providing parton showers with
matrix-element corrections we still only get the leading-order normalization, 
because in the initial-state cascade we only include leading 
logs and not the full one-loop virtual contributions.

In Section 2 we review the basis of the HERWIG parton shower algorithm for
the initial-state radiation in hadronic collisions. In Sections 3 and 4 we
discuss the hard and soft matrix-element corrections to vector boson
production.
In Section 5 we plot some
relevant phenomenological distributions at the centre-of-mass energy of the
Tevatron and of the LHC using the new version of HERWIG\@. We compare
our results with previous versions of HERWIG, resummed calculations and
experimental data. Finally, in Section 6 we discuss our results and make
some concluding comments.

\section{The parton shower algorithm}

The production of a vector boson $V$ in hadronic collisions
is given at lowest order by the elementary parton-level process
$q\bar q\to V$. In the following, we shall assume that the vector boson decays
into a lepton pair (Drell--Yan interactions).
The first-order tree-level corrections to such a process are given
by the processes $q\bar q \to V g$ and $q g\to V q$
($\bar q g\to V\bar q$), where the initial-state partons can come from
either incoming hadron.

A possible method to implement the initial-state parton shower in a
probabilistic way is the Altarelli--Parisi approach, in which
the initial energy scale $Q_0$ is increased up to the probed value $Q$
and all the effect of the emitted partons is integrated out.

On the contrary, standard Monte Carlo programs [\ref{sjostrand},\ref{marweb}]
explicitly keep track of the accompanying radiation, by implementing the
so-called `backward evolution' in which the hard
scale is reduced away from the hard vertex, tracing the hard scattering
partons back into the original incoming hadrons and
explicitly generating the distribution of emitted partons.
In the leading infrared approximation, the probability of the emission
of an additional parton from a parton $i$ is given by the general result for
the radiation of a soft/collinear parton:
\begin{equation}
  \label{elementary}
  dP={{dq_i^2}\over{q_i^2}}\;
  {{\as\left(\frac{1-z_i}{z_i}q_i\right)}\over {2\pi}}\;
  P_{ab}(z_i)\; dz_i\;
  {{\Delta_{S,a}(q^2_{i\mathrm{max}},q_c^2)}\over{\Delta_{S,a}(q_i^2,q_c^2)}}\;
{{x_i/z_i}\over x_i}\;{{f_b(x_i/z_i,q_i^2)}\over {f_a(x_i,q_i^2)}}.
\end{equation}

The HERWIG parton shower is ordered according to the variable
$q_i^2=E^2\xi_i$, where $E$ is the energy of the parton that split
and $\xi_i={{p_h\cdot p_i}\over{E_h E_i}}$, where $p_i$ is the 
four-momentum of the emitted parton; $p_h$ is a lightlike vector with
momentum component parallel to the incoming hadron; $E_h$ and $E_i$ are
the energy components of $p_h$ and $p_i$; and
$z_i$ is the energy fraction of the outgoing space-like
parton (which goes on to participate in the hard process) with respect
to the incoming one (i.e.~$z_i=1-E_i/E$).
In the approximation of massless partons, we have
$\xi_i=1-\cos\theta$, where $\theta$ is the emission angle from the
incoming hadron direction. When all emission is soft, the energy of the emitted
partons is negligible ($E_i\ll E$), therefore ordering according to $q_i^2$
corresponds to angular ordering; when the emission is hard, the energy of
the radiated parton is similar to that of the splitting parton, so
$q_i^2$ ordering is equivalent to transverse momentum ordering.
In (\ref{elementary}) $f_a(x_i,q_i^2)$ is the parton distribution function
for the partons of type $a$ in the initial-state hadron, $x_i$ being the
parton energy fraction.
At each step of the backward evolution a parton of type $a$, a quark for
example, can evolve back to any other type of parton $b$, in this case
either a quark of the same flavour or a gluon, having a higher value of
$x_i$.

The quantity $\Delta_{S,a}(q_i^2,q_c^2)$ is the Sudakov form factor,
resulting from the leading-logarithmic resummation and
representing the probability that no resolvable radiation is
emitted from a parton of type $a$ whose upper limit on emission is
$q_i^2$, with $q_c^2$ being, in the case of HERWIG, a cutoff on
transverse momentum.
This cutoff implies a minimum value of the evolution scale $q_i^2$ that
can be reached, $q_i^2>4q_c^2$, but in practice this is smaller than the
smallest scale at which most standard parton distribution function sets
are reliable, so an additional cutoff on $q_i^2$ has to be
applied.
The ratio of form factors appearing in Eq.~(\ref{elementary}) represents the
probability that the emission considered is the first, i.e.\ the one with the
highest value of $q_i^2$. In terms of Feynman diagrams, the Sudakov form factor
sums up all-order virtual and unresolved contributions.
$P_{ab}(z_i)$ is the Altarelli-Parisi splitting function for a parton of
type $b$ to evolve to one of type $a$ with momentum fraction $z_i$.
$\as$ is the strong coupling, evaluated at a scale of order the
transverse momentum of the emitted parton, which sums large higher order
corrections. This, together with the angular ordering condition, makes
HERWIG accurate to next-to-leading order at large $x$ [\ref{CMW}].

The definition of the variables $q_i^2$ and $z_i$ is not Lorentz-invariant, but
it is frame-dependent.
Colour coherence implies that for any pair of colour-connected partons
$i$ and $j$ the maximum values of the $q$ variables are related by
$q_{i\mathrm{max}}q_{j\mathrm{max}}=p_i\cdot p_j$.  Therefore one is free to
choose the frame in which to define the initiating values $q_{i\mathrm{max}}$
and $q_{j\mathrm{max}}$, with the only prescription being that their product
must equal $p_i\cdot p_j$. The subsequent emissions are then ordered in
$q_i^2$.
For vector boson production, as in most cases, symmetric limits are fixed by
HERWIG, i.e.\ $q_{i\mathrm{max}}^2=q_{j\mathrm{max}}^2=p_i\cdot p_j$
and the energy of the parton which initiates the cascade is set
to $E=q_{\mathrm{max}}=\sqrt{p_i\cdot p_j}$.
Ordering according to $q_i^2$ therefore dictates $\xi_i<z_i^2$.

After we generate the initial-state shower, the original partons are not
on their mass-shell anymore, so their energy and momentum cannot be conserved.
Energy-momentum conservation is then achieved by applying a separate
boost to each
jet along its own direction. As a result of this, the jet momenta are no
longer equal to the parton ones, but energy and momentum are globally
conserved and the vector boson acquires a transverse momentum
from the recoil against the emitted partons.
Since the mass shift becomes negligible in the soft and collinear
limits, the precise details of this kinematic reshuffling are not fixed
{\it a priori}, but are free choices of the model.

Once the backward evolution has terminated, a model to reconstruct the original
hadron is required. In HERWIG, if the backward evolution has not resulted in a
valence quark, additional non-perturbative parton emission is generated to
evolve back to a valence quark. Such a valence quark has a Gaussian
distribution with respect to the non-perturbative intrinsic transverse momentum
in the hadron, with a width that is an adjustable parameter of the
model.  In the following, when discussing the phenomenological implications of
our work, we shall consider both HERWIG's default value of zero, and an
increased value of 1 GeV, bracketing the reasonable range of
non-perturbative effects.

The algorithm so far discussed is reliable only in the soft or collinear
limits and, since it only describes radiation for $\xi_i<z_i^2$, there are
regions of the phase space that are
completely empty (`dead zones'). The radiation in such regions, according to
the full matrix element, should be suppressed, but not completely absent as
happens in HERWIG\@.
We therefore need to improve the HERWIG model by implementing
matrix-element corrections.
As usual[\ref{sey1}--\ref{corcella}], this method works in two steps: we
populate the missing phase space region by
generating radiation according to a distribution obtained from the
first-order matrix-element calculation (`hard corrections'); we correct
the algorithm in the already-populated region using the matrix-element
probability distribution whenever an emission is capable of being the
`hardest so far' (`soft corrections').

\section{Hard Corrections}

In order to implement the hard and soft matrix-element corrections to
simulations of the initial-state radiation in Drell--Yan processes, we firstly
have to relate the HERWIG variables $\xi$ and $z$ to the kinematic ones we
use in the matrix-element calculation.
For the process $q(p_1)\bar q(p_2)\to g(p_3) V(q)$ we parametrize the phase
space according to the Mandelstam variables $\hat s = (p_1+p_2)^2$,
$\hat t=(p_1-p_3)^2$ and $\hat u=(p_2-p_3)^2$.
Throughout this paper, we neglect the parton masses, so we have
$\hat s+\hat t+\hat u=m_V^2$.

The phase space limits, in terms of the variables $\hat s$ and $\hat t$, are:
\begin{eqnarray}
  \hspace{-2cm}
   m_V^2 \;\;<&\hat s&<\;\; s,
  \\\hspace{-2cm}
  m_V^2-\hat s \;\;<&\hat t&<\;\; 0,
\end{eqnarray}
where $s$ is the squared energy in the centre-of-mass frame.
Note that the point $\hat s=m_V^2$ corresponds to the soft singularity,
and the lines $\hat t=0$ and $\hat t=m_V^2-\hat s$ to collinear
emission.

In order to relate $\hat s$ and $\hat t$ to $\xi$ and $z$ we use the property
that the mass $m$ and the transverse momentum $p_t$ of the $q$--$g$
($\bar q$--$g$) jets are conserved in the showering frame.
In doing this, we observe that, in the approximation of massless partons,
the energy of the annihilating $q\bar q$ pair which produces the vector boson
$V$ is equal to $E' =\sqrt{p_q\cdot p_{\bar q}}=\sqrt{m_V^2/2}$.

In terms of the showering variables, we obtain:
\begin{eqnarray}
  m^2   &=& -{{1-z}\over {z^2}}\;\xi\;m_V^2,\\
  p_t^2 &=& {{(1-z)^2}\over{2 z^2}}\xi\;(2-\xi)\;m_V^2,
\end{eqnarray}
and in terms of the matrix-element variables:
\begin{eqnarray}
  m^2   &=& \hat t,\\
  p_t^2 &=& {{\hat u\hat t}\over {\hat s}}.
\end{eqnarray}
Combining them we get the following equations:
\begin{eqnarray}
  z &=&{{m_V^2}\over\hat t}+\sqrt{\left({{m_V^2}\over\hat t}\right)^2
  -{{2m_V^2}\over{\hat s\hat t}}(m_V^2-\hat t)}\;\;,\\
  \xi &=& -2\;{{{{m_V^2}\over\hat t}-{{m_V^2-\hat t}\over{\hat s}}\;
+\;m_V^2\;\sqrt{{1\over{\hat t^2}}-{2\over{\hat s\hat t}}+
{2\over {m_V^2\hat s}}}}\over {1-{{m_V^2}\over {\hat t}}-
m_V^2\;\sqrt{{1\over {\hat t^2}}-{2\over{\hat s\hat t}}+
{2\over {m_V^2\hat s}}}}}\;.
\end{eqnarray}

The region where HERWIG does not allow gluon radiation can be derived by
solving the equation $\xi>z^2$:
\begin{eqnarray}
  \hat s_{\mathrm{min}}\;<&\hat s&<\;s \\
  \hat t_{\mathrm{min}}\;<&\hat t&<\;\hat t_{\mathrm{max}},
\end{eqnarray}
where $\hat t_{\mathrm{max}}$ can be obtained by solving the equation
\begin{equation}
\hat t^2+3m_V^2\hat t+2m_V^4\left(1-{{m_V^2}\over \hat s}\right)
=0,
\end{equation}
or:
\begin{equation}
\hat t_{\mathrm{max}}=
  -{{m_V^2}\over 2}\;\left(3-\sqrt{1+8m_V^2/\hat s}\right)\;.
\end{equation}
It is straightforward to write $\hat t_{\mathrm{min}}$ as
\begin{equation}
\hat t_{\mathrm{min}}=m_V^2-\hat s- \hat t_{\mathrm{max}}\;,
\end{equation}
while $\hat s_{\mathrm{min}}$ can be determined by the condition
$\hat t_{\mathrm{min}}(\hat s)<\hat t_{\mathrm{max}}(\hat s)$:
\begin{equation}
\hat s_{\mathrm{min}}={{m_V^2}\over 2} \left( 7-\sqrt{17}\right).
\end{equation}

\begin{figure}[t]
\centerline{\resizebox{0.65\textwidth}{!}{\includegraphics{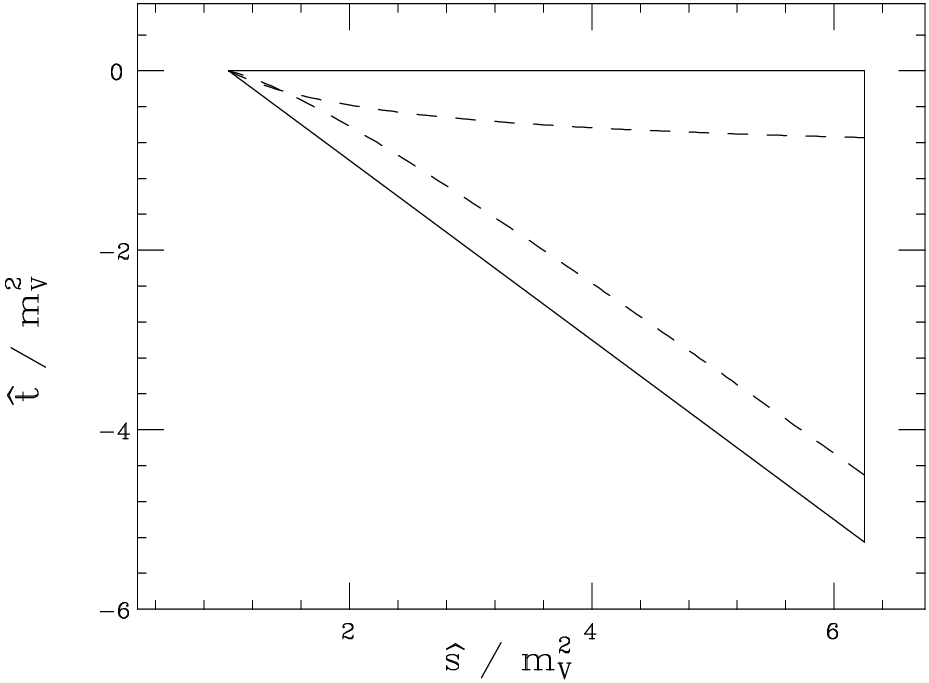}}}
\caption{The phase space for additional emission in Drell--Yan
  production with $s/m_V^2=6.25$, showing the kinematic limits (solid)
  and HERWIG's parton shower limits (dashed).}
\label{fig:phase}
\end{figure}

In Fig.~\ref{fig:phase} we plot the total phase space and HERWIG's limits for
a vector boson mass of $m_V=80$~GeV and a centre-of-mass energy of
$\sqrt{s}=200$~GeV, in terms of the normalized variables
${{\hat s}/{m_V^2}}$ and ${{\hat t}/{m_V^2}}$.
We can see that, as in cases [\ref{sey2}] and [\ref{sey3}]
and differently from [\ref{corcella}], the soft and the collinear singularities
are well inside the HERWIG phase space: we also have an overlapping region,
corresponding to a kinematic configuration in $\hat s$ and $\hat t$ where
radiation can come from either parton.
Note that the only dependence on the external physical parameters is the
position
of the edge at large $\hat s$, i.e.\ $\hat s_{\mathrm{max}}=s$, while the
values of $\hat s_{\mathrm{min}}/m_V^2$ and the limits in $\hat t/m_V^2$
are independent
of the centre-of-mass energy and of other kinematic conditions like the
vector boson rapidity.

Once we have the total and HERWIG phase space limits, in order to implement
matrix-element corrections, we have to apply the exact differential
cross section in the dead zone.
In [\ref{sey4}] a general prescription is given to allow first-order
corrections to quark scattering and annihilation processes once a generator of
the lowest order process is available.
For the Drell--Yan case, assuming that the virtuality and the
rapidity of the produced vector boson are fixed by the Born process
$q\bar q\to V$, the first-order differential cross section $d\sigma$ is
proportional to the
parton-level lowest order $\sigma_0$ according to the relation:
\begin{equation}
\label{factorization}
d^2\sigma = \sigma_0{{f_{q/1}(\chi_1)f_{\bar q/2}(\chi_2)}\over
{f_{q/1}(\eta_1)f_{\bar q/2}(\eta_2)}}{{C_F\;\as}\over {2\pi}}
{{d\hat s\; d\hat t}\over{\hat s^2\hat t\hat u}}
\left[(m_V^2-\hat u)^2+(m_V^2-\hat t)^2\right].
\end{equation}
In the above equation $f_{q/1}(\chi_1)$ and $f_{\bar q/2}(\chi_2)$ are the
parton distribution functions of the scattering partons inside the incoming
hadrons 1 and 2 for energy fractions $\chi_1$ and $\chi_2$ in the process
$q\bar q\to V g$, while $f_{q/1}(\eta_1)$ and $f_{\bar q/2}(\eta_2)$ refer
to the Born process and cancel off the factors that are already in $\sigma_0$.
The assumption that the rapidity and mass of the vector boson are the same as
in the process $q\bar q\to V$ allows us to recover the Born result in the
limit of an extremely soft gluon radiation.

As stated here, Eq.~(\ref{factorization}) is a trivial rewriting of the
first-order differential cross section, but the main point of
[\ref{sey4}] is that if the azimuth of the emitted gluon is generated in
the right way, Eq.~(\ref{factorization}) correctly describes the full
process including the vector boson decay.  Thus, to implement our
matrix-element corrections, we do not need to know anything about the final
state of the vector boson~-- its properties are correctly inherited
from the Born process.

In a similar way we deal with the Compton process
$q(p_1) g(p_3)\to q(p_2) V(q)$.
We define the Mandelstam variables $\hat s=(p_1+p_3)^2$,
$\hat t=(p_3-p_2)^2$ and $\hat u=(p_1-p_2)^2$, and we find
the same expressions for the variables $\xi$ and $z$ and for the phase space
limits.
We obtain for the differential cross section [\ref{sey4}]:
\begin{equation} \label{qg}
d^2\sigma = -\sigma_0 {{f_{q/1}(\chi_1)f_{g/2}(\chi_2)}\over
{f_{q/1}(\eta_1)f_{\bar q/2}(\eta_2)}}{{T_R\as}\over {2\pi}}
{{d\hat s\; d\hat t}\over{\hat s^3\hat t}}
\left[(m_V^2-\hat t)^2+(m_V^2-\hat s)^2\right].
\end{equation}

Extending this formula to processes where we have an incoming antiquark or
where the gluon belongs to hadron~1 is straightforward.
We then generate events according to the above distributions in the dead zone
using standard techniques.

When applying the hard corrections, in principle one should also
implement the form factor, but, since we are quite far from the soft and
collinear singularities, it is actually not important and we shall
neglect it in the following.  This is justified by the fact that the
total fraction of events that receive an emission from the hard
correction is small.  For example for $W$ production it is 3.9\% at the
Tevatron and 9.2\% at the LHC\@.
Also, the fact that such fractions are quite small allows us to neglect
multiple
emissions in the dead zone and makes the use of the exact first-order result
reliable.

Among these events, the fraction of $q\bar q'\to W g$ processes is 53.5\%
at the Tevatron and 24.5\% at the LHC\@.  The reason for the differences
between the two machines can be understood in terms of the parton
distribution functions.  The gluon density inside the protons is higher
when the colliding energy is increased because $x$ is decreased;
moreover at the LHC we have $pp$ interactions instead of $p\bar p$,
therefore a $q\bar q'$ annihilation requires an antiquark $\bar q'$ to be
taken from the `sea'.

The equivalent numbers for $Z$ production are essentially identical.

\section{Soft Corrections}

According to [\ref{sey1}], we should also correct the emission in the region
that is already populated by HERWIG using the exact first-order calculation
for every emission {\em that is the hardest so far}.
This can be performed by multiplying the parton shower distribution by
a factor that is equal to the ratio of
HERWIG's differential distribution to the matrix-element one.  The only
non-trivial part of this is in calculating the Jacobian factor
$J(\hat s,\hat t;z,\xi)$ of the transformation $(z,\xi)\to (\hat s,\hat
t)$. HERWIG's cross section is then given by
\begin{equation}
  {{d^2\sigma}\over{d\hat s d\hat t}}={{d^2\sigma}\over{dz d\xi}} \;
  J(\hat s,\hat t;z,\xi),
\end{equation}
where ${{d^2\sigma}/{dz d\xi}}$ is given by the elementary emission
probability given in Eq.~(\ref{elementary}).  The Jacobian factor
$J$ can be simply calculated from the relations given earlier:
\begin{equation}
  J(\hat s,\hat t;z,\xi)=\frac{\hat t(m_V^2-\hat t)}{\hat s^2}\;
  \frac{z^5}{m_V^4\xi(1-z)^2\left(z+\xi(1-z)\right)}\;.
\end{equation}

At this point we are able to make some comparisons with the approach that is
followed in [\ref{miu}] where matrix-element corrections are added to
the PYTHIA simulation of vector boson production. The parton shower
probability distribution is applied over the whole phase space 
(in its older versions
PYTHIA had a cutoff on the virtuality $k^2$ of the hard scattering parton that
was constrained to be $k^2<m_W^2$ in order to avoid double counting) and the
exact matrix-element correction is applied only to the branching that is
closest to the hard vertex.
Unlike [\ref{miu}], we have complementary phase space regions where we
apply either
the parton shower distribution (\ref{elementary}) or the exact matrix-element
ones (\ref{factorization},\ref{qg}), while in the parton shower region
($\xi<z^2$) we use the exact
amplitude to generate the hardest emission so far instead of just the first
emission.
Correcting only the first emission can lead to problems due to the
implementation of the Sudakov form factor whenever a subsequent harder
emission occurs, as we would get the unphysical result that the probability of
the hard emission would depend on the infrared cutoff that appears in the
expression of the form factor.
See [\ref{sey1}] for more details on this point.

\section{Results}

Having implemented the hard and soft matrix-element corrections to the
initial-state radiation in Drell--Yan interactions, we wish to
investigate the impact they
have on relevant phenomenological observables that can be measured at the
Tevatron and in future at the LHC\@.
In the following, we shall mostly concentrate on $W$ production although
$\gamma^*$ and $Z$ events are treated in exactly the same way.

\subsection{Vector boson transverse momentum}

A particularly significant phenomenological quantity is the transverse momentum
of the $W$ to the beam axis, which has been object of many theoretical and
experimental analyses.
In the soft/collinear limit, the transverse momentum of the $W$ is
constrained to be $q_T<m_W$, since in the hard process $q\bar q'\to W$ the $W$
is produced with no transverse momentum and
it can acquire some $q_T$ only as a result of the initial-state
parton showering.
When the emission is generated according to the exact matrix element of
processes like $q\bar q' \to W g$, $q g\to q' W$ or $\bar q g\to \bar q' W$,
the $W$ produced in the hard process is allowed to have a non-zero $q_T$ and
events with $q_T>m_W$ are expected.

\begin{figure}[t]
\centerline{\resizebox{0.65\textwidth}{!}{\includegraphics{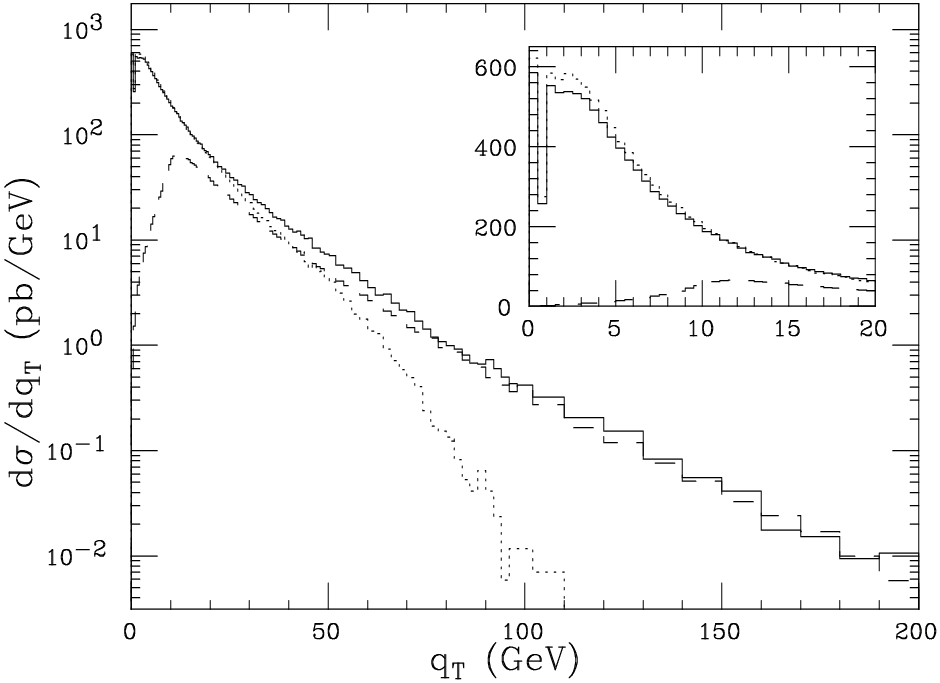}}}
\caption{The $W$ $q_T$ distribution at the Tevatron, according to HERWIG
  without (dotted) and with (solid) matrix-element corrections.  Also
  shown (dashed) is the `$W+$~jet' process with a cutoff of
  10~GeV.}
\label{fig:WqtTev}
\end{figure}

In Fig.~\ref{fig:WqtTev} we compare the differential cross sections with
respect to the $W$
$q_T$ for $p\bar p$ collisions at the Tevatron energy\footnote{The next
  Tevatron run will be at the slightly higher energy of 2~TeV.  For the
  sake of comparison with existing data we use 1.8~TeV, but the results
  would not be qualitatively different for 2~TeV.}, $\sqrt{s}=1.8$~TeV,
obtained using HERWIG 5.9, the latest public version,
with 6.1, the new version in progress where we include for
the first time matrix-element
corrections to the initial-state parton shower in vector boson production.
We set an intrinsic transverse momentum equal to zero and
use the MRS (R1) parton distribution functions [\ref{mrs}].
We can see from the plots that the impact of matrix-element corrections is
negligible at low values of $q_T$ but it is quite relevant at high $q_T$,
where we have many more events with respect to the 5.9 version. Above some
value of $q_T$ HERWIG 5.9 does not generate events anymore, while the 6.1
version still gives a non-zero differential cross section
thanks to the events generated via the exact hard matrix element.
As in $e^+e^-$ annihilation[\ref{sey2}] and DIS[\ref{sey3}], it is
actually the hard corrections that have a marked impact
on our distributions, while the effect of the soft ones is quite negligible.

It is also interesting to plot the $q_T$ spectrum obtained running the
`$W$ + jets'~process
of HERWIG forcing the produced $W$ to decay leptonically.
This generates the hard
process $q\bar q'\to W g$ (or the equivalent ones with an initial-state gluon)
for all events. As this matrix element diverges when the transverse
momentum of the $W$ approaches zero, HERWIG applies a user-defined cut
on the $q_T$ generated in the hard process, which we set to 10~GeV. In
our plot, this does not appear as a sharp cutoff
since the $W$ gets some recoil momentum due to the initial-state parton shower
which can increase or decrease its transverse momentum.
If, on the contrary, we had plotted the $q_T$ of the $W$ generated in the hard
$2\to 2$ process, we
would have got a sharp peak at $q_T=$~10~GeV.
The agreement we find between the simulations of Drell--Yan
processes provided with matrix-element corrections and the `$W$ + jet' events
for large $q_T$ reassures us that the implementation of the hard
corrections is reliable.

We also see that the HERWIG distributions for Drell--Yan processes show
a sharp peak in the first bin, which includes the value $q_T=0$: it
corresponds to a fraction of events with no initial-state radiation and
so $W$ bosons produced with zero transverse momentum.
With any fixed infrared cutoff value, one expects a non-zero (though
exponentially suppressed) fraction of events to give no resolvable
radiation.  These would normally be smeared out by non-perturbative
effects like the intrinsic transverse momentum, as we shall see in later
plots, but since we set the width of its distribution to zero by
default, all such events appear in the lowest $q_T$ bin.
This is actually a technical deficiency of the Monte Carlo simulation and not
a detectable physical effect.

\begin{figure}[t]
\centerline{\resizebox{0.65\textwidth}{!}{\includegraphics{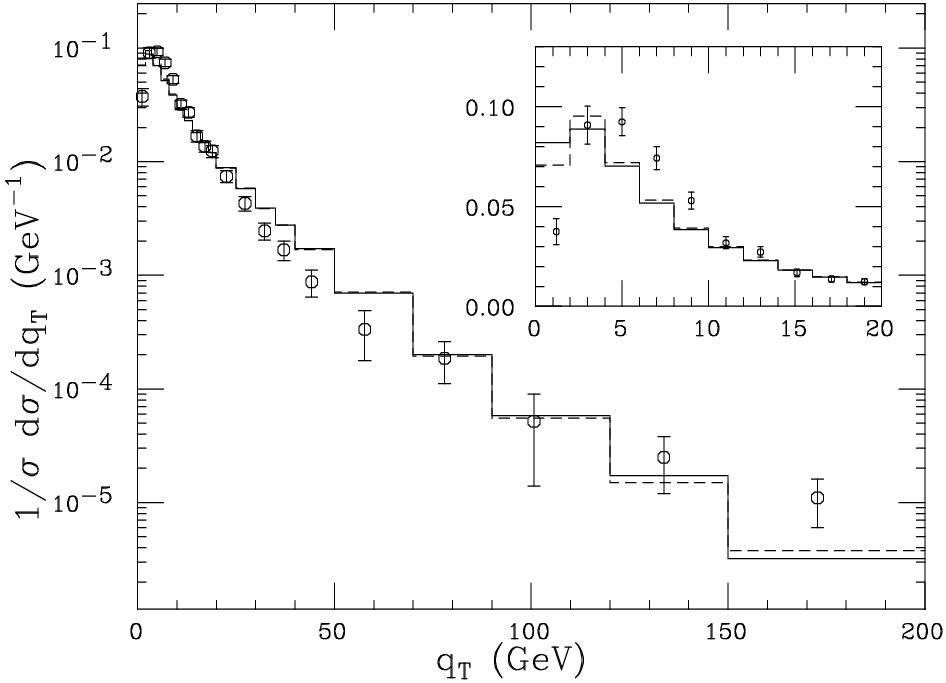}}}
\caption{The $W$ $q_T$ distribution data from D\O, in comparison with HERWIG
  with matrix-element corrections but without detector corrections.  The
  solid line has zero intrinsic transverse momentum while the dashed one
  has an r.m.s.~$p_t$ of 1~GeV.}
\label{fig:WqtD0}
\end{figure}

The D\O\ collaboration recently published data on the transverse momentum
distribution of $W$ bosons at the Tevatron[\ref{d0}].  In
Fig.~\ref{fig:WqtD0} we compare the HERWIG predictions with it.
\begin{figure}[t]
\centerline{\resizebox{0.65\textwidth}{!}{\includegraphics{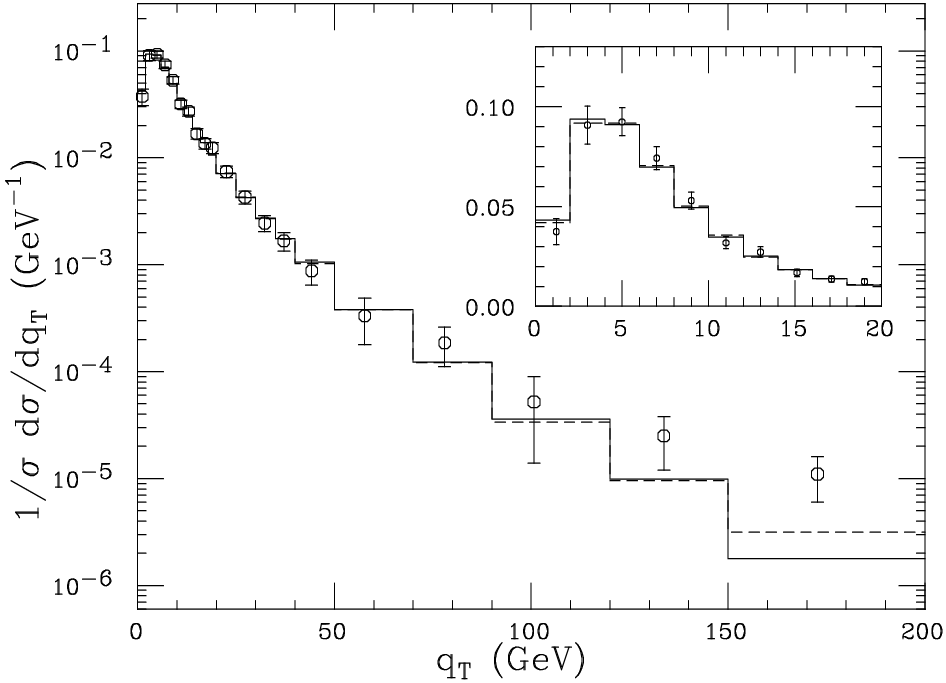}}}
\caption{As Fig.~\ref{fig:WqtD0} but with the HERWIG results corrected
  for detector effects.}
\label{fig:WqtD0_corrected}
\end{figure}
In order to contribute to the investigation of possible effects of
non-perturbative physics, we also run HERWIG setting an intrinsic partonic
transverse momentum equal to 1~GeV, which we consider to be the maximum
reasonable value.

We see that the data has a significantly broader distribution at small
$q_T$ than HERWIG without intrinsic transverse momentum and that
increasing the r.m.s.~$p_t$ to 1~GeV is nowhere near enough to account
for this.  Furthermore the description of the data in the intermediate
$q_T$ range, 30--70~GeV, is also rather poor.  The intrinsic $p_t$ does
not affect the predictions significantly for $q_T$ values above about
10~GeV, so there is no obvious way to improve the fit for intermediate
values.

However, these effects are actually because the D\O\ data are
uncorrected for detector effects.  We have run HERWIG through D\O's fast
simulation program, \texttt{CMS}[\ref{CMS}], and show results in
Fig.~\ref{fig:WqtD0_corrected}.  We see that HERWIG now describes the
data rather well.  The detector smearing is so strong at low $q_T$ that
the additional smearing produced by the intrinsic transverse
momentum becomes irrelevant.

At this point it is worthwhile commenting on the results shown in
Refs.~[\ref{miu},\ref{mrenna}].  Both compare generator-level
results with the D\O\ data.
Ref.~[\ref{mrenna}]'s actually look rather similar to our
generator-level results, so it is likely that after applying detector
corrections they will describe the data as well as HERWIG\@.
Ref.~[\ref{miu}] found good agreement with the D\O\ data, but only after
increasing the intrinsic transverse momentum to 4~GeV.  It seems likely
that this accounts for the smearing at low $q_T$ and would not be
necessary after including detector smearing.  However, in the
intermediate $q_T$ range the results of [\ref{miu}] are significantly
lower than
HERWIG\@.  Since we find a detector correction of around a factor of two
in this region, it would be very interesting to see the results of
[\ref{miu}] at detector level to see whether they are still able to fit
the data.

\begin{figure}[t]
\centerline{\resizebox{0.65\textwidth}{!}{\includegraphics{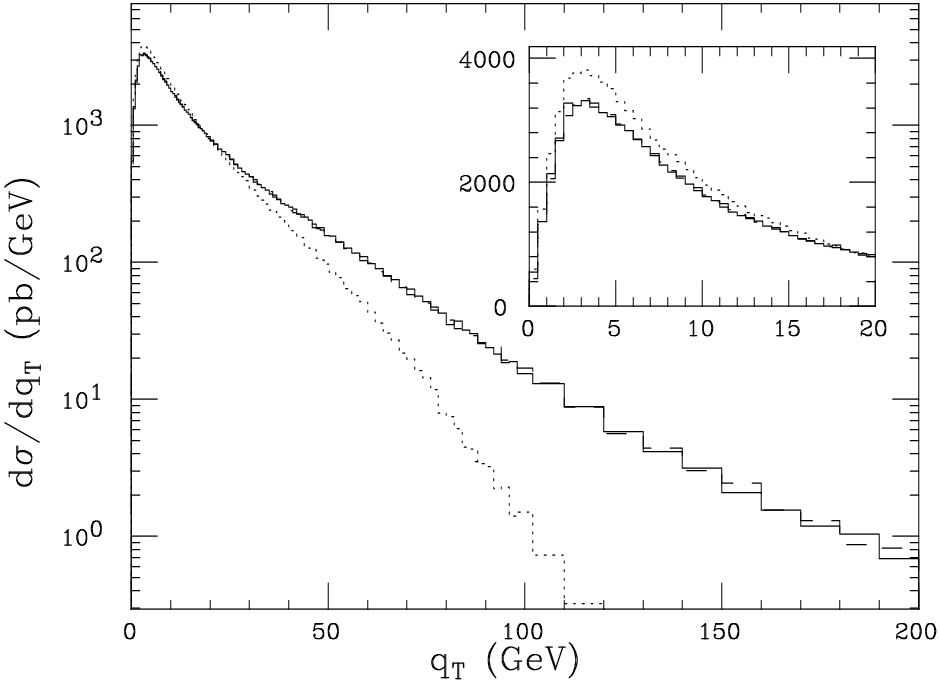}}}
\caption{The $W$ $q_T$ distribution at the LHC, according to HERWIG
  without (dotted) and with (solid and dashed) matrix-element
  corrections, with zero intrinsic $p_t$ (solid) and an r.m.s.~$p_t$ of
  1~GeV (dashed).}
\label{fig:WqtLHC}
\end{figure}

In Fig.~\ref{fig:WqtLHC} we show the $q_T$ distributions for $pp$
collisions at the energy of
the LHC, $\sqrt{s}=14$~TeV, and find that the impact of the corrections
is even bigger once the energy is increased.
Unlike in the Tevatron transverse momentum distributions, we do not have
the previously-mentioned sharp peak at $q_T=0$: this is because at the
LHC we have $pp$ interactions and the protons do not have valence
antiquarks, while in order to produce a $W$ we do need a $q\bar q'$ hard
scattering.  As a result, the backward evolution has to produce at least
one splitting, which always gives the $W$ itself some transverse momentum.
From the window in the top-right corner, we also see that at very low $q_T$
the uncorrected version, 5.9, has a few percent more events than 6.1,
particularly in the case of the LHC\@.
As we said in the introduction, although we have matched to the tree-level
NLO matrix elements, we still get the LO normalization, therefore the total
cross sections obtained from versions 5.9 and 6.1 are the same.
Since at the energy of the LHC we are generating a higher fraction of events 
at large $q_T$ via the exact matrix-element distribution, it is reasonable 
that this enhancement is partially compensated by a slight suppression in the 
low $q_T$ region.

In the region of low $q_T$, it is worthwhile comparing the HERWIG
distributions with some resummed calculations that are available in the
literature.  All these calculations are based on the approach suggested
in [\ref{ddt}] where the differential cross section with respect to the
vector boson $q_T$ is expressed as the resummation of logarithms $l=\log
(m_V^2/q_T^2)$ to all orders in $\as$.
Two conflicting nomenclatures are used in the literature to denote which
logarithms are summed: in the differential cross section, at each order
in $\as$ the largest term is $\sim1/q_T^2\,\as^n\,l^{2n-1}$,
which are sometimes known as the leading logarithms,
$\sim1/q_T^2\,\as^n\,l^{2n-2}$ being known as the next-to-leading
logarithms, and so on. According to this classification, the results in 
[\ref{ellis2}] and [\ref{stirling}] are NNLL and NNNLL respectively.

However, these logarithms `exponentiate',
allowing the differential cross section to be written in terms of the
exponential (the `form factor') of a series in $\as$ whose largest
term is $\sim\as^n\,l^{n+1}$, which are also sometimes known as the
leading logarithms, $\sim\as^n\,l^n$ being known as the
next-to-leading logarithms, and so on.
In [\ref{nason}] all NLL terms according to this nomenclature are summed in 
the form factor, which is evaluated
either in the impact parameter $b$-space or in the $q_T$-space.
The non-perturbative contribution is taken into account in the $b$-space
formalism following the
general ideas in [\ref{ly}] and setting Gaussian functions in the impact
parameter $b$ to quantify these effects.

\begin{figure}[t]
\centerline{\resizebox{0.65\textwidth}{!}{\includegraphics{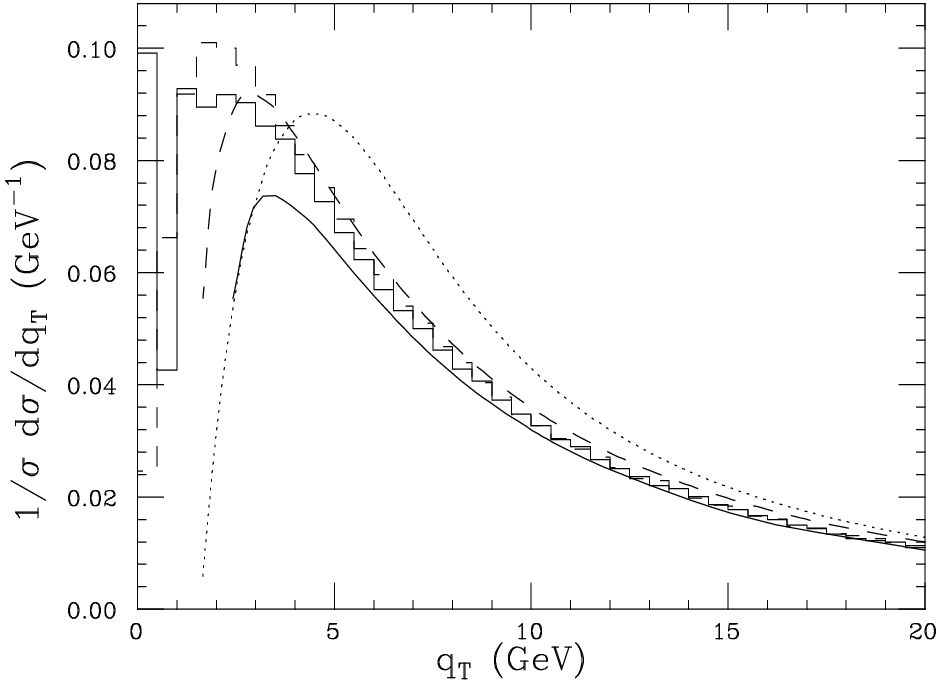}}}
\caption{The $W$ $q_T$ distribution at the Tevatron, according to HERWIG
  with matrix-element corrections, with zero intrinsic $p_t$ (solid
  histogram) and an r.m.s.~$p_t$ of 1~GeV (dashed histogram), compared
  with the resummed results of [\ref{nason}] in $q_T$-space (solid) and
  in $b$-space (dotted) and of [\ref{ellis2}] in $q_T$-space (dashed).}
\label{fig:resum1}
\end{figure}
In Fig.~\ref{fig:resum1} we compare the HERWIG 6.1 differential cross
section\footnote{As the resummed calculations deal with a fixed value of
  $m_W$, Fig.~\ref{fig:resum1} is obtained running HERWIG with a
  vanishingly small $W$ width, so $m_W\simeq 80.4$~GeV, its default
  value.  $W$ width effects are nevertheless fully included in HERWIG
  and in the other plots we show. At low $q_T$ this assumption does not
  change the results dramatically, at high $q_T$ the effect of the $W$
  width is important as it allows values of $q_T$ larger that the
  default $W$ mass even in the parton shower approximation, otherwise
  they could come only via the exact matrix-element generated events.}
with the ones obtained from these approaches [\ref{ellis1}--\ref{nason}]
all normalized to the corresponding total cross section.
HERWIG clearly lies well within the range of the resummed approaches,
except at very small $q_T$ where they become unreliable.
To be more precise, the agreement is better between HERWIG and the two
resummations in the $q_T$-space, but even the $b$-space result is not too far
from the Monte Carlo distribution.

If we now wish to compare the HERWIG simulation with matrix-element
corrections to the resummed calculations even for larger values of $q_T$ we
need to match the latter with the exact ${\cal O}(\as)$ results
to make them reliable there.
This has already been done in the literature within the approach of
[\ref{ellis1},\ref{ellis2}], while in [\ref{nason}] the
analysis is limited to the low $q_T$ regime.

Many prescriptions exist concerning how to perform such a matching.
Ours is to simply add the exact matrix-element cross sections for the parton
level processes $q\bar q'\to Wg$ and $q(\bar q) g\to W q'(\bar q')$, already
calculated in (\ref{factorization}) and (\ref{qg}),
to the resummed expressions and, in order to avoid double counting,
subtract off the terms they have in common.  It is straightforward to
show that these are simply those terms in the exact ${\cal O}(\as)$
result that do not vanish as $q_T\to0$.
This prescription works fine if the resulting distribution is continuous at the
point $q_T=m_W$, which means that the resummation and the low $q_T$
${\cal O} (\as)$ result exactly compensate each other and only the exact
`hard' matrix-element contribution survives.

As discussed in [\ref{ellis1},\ref{ellis2}], it is not trivial to
implement such a matching: the authors in fact do not succeed in obtaining a
continuous distribution at the crucial matching point $q_T=m_W$, but
rather a step of size $\sim\as^2$ was found.  This comes about
because the derivative of the Sudakov exponent is not required to go
smoothly to zero at that point.

We independently implement the matching for all the resummed calculations with
which we wish to compare the HERWIG results and we find that the matching works
well only for the $q_T$-space resummation performed in [\ref{nason}].
For the $b$-space and the approaches in [\ref{ellis2}] we
do indeed find a step at $q_T=m_W$.  In fact even for the $q_T$-space
method of [\ref{nason}] we find a `kink' at $q_T=m_W$, i.e.\ although
the curve is continuous, its derivative changes discontinuously there,
albeit by an amount that is too small to notice on the figure.

\begin{figure}[t]
\centerline{\resizebox{0.65\textwidth}{!}{\includegraphics{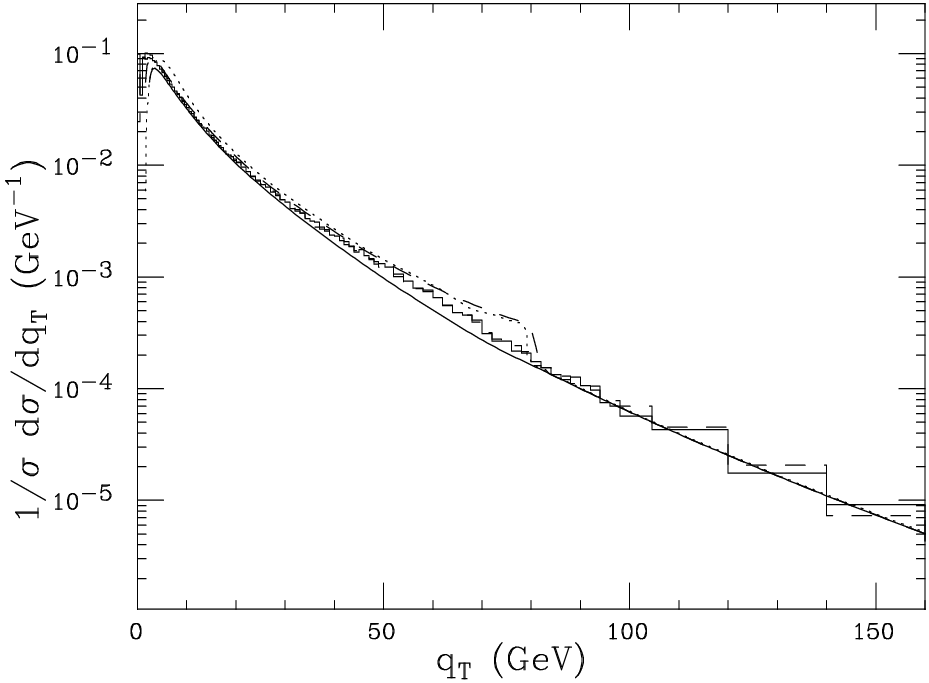}}}
\caption{As Fig.~\ref{fig:resum1}, but with the resummed results matched
  with the exact ${\cal O}(\as)$ result.}
\label{fig:resum2}
\end{figure}

In Fig.~\ref{fig:resum2} we compare the HERWIG 6.1 distributions with
[\ref{ellis1}--\ref{nason}] after the matching over the whole $q_T$ spectrum.
We see that for the resummed distribution that is well matched and
continuous for $q_T=m_W$ the agreement with HERWIG is pretty
good everywhere.  We do have slight discrepancies for medium values of
$q_T$, but they are well within the range that could be expected from
the differences between the approaches followed by the Monte Carlo
program and the calculation which keeps all the 
next-to-leading logarithms in the form factor.

\begin{figure}[t]
\vspace*{-1cm}
\centerline{\resizebox{0.65\textwidth}{!}{\includegraphics{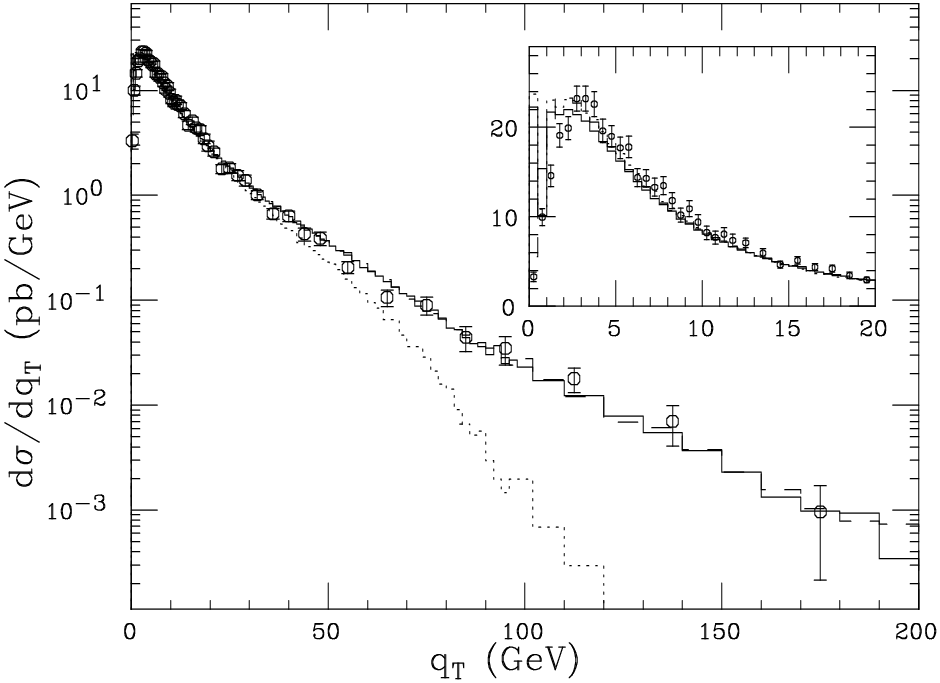}}}
\caption{The preliminary $Z$ $q_T$ distribution data from CDF, in
  comparison with HERWIG without (dotted) and with (solid and dashed)
  matrix-element corrections.  The solid and dotted lines have zero
  intrinsic transverse momentum while the dashed one has an r.m.s.~$p_t$
  of 1~GeV.}
\label{fig:ZqtCDF}
\end{figure}

While the plots shown so far refer to $W$ production, it is also worth
comparing with some recent preliminary CDF data on $Z$ production [\ref{cdf}].
In Fig.~\ref{fig:ZqtCDF}, we have the CDF distribution with respect to
the transverse
momentum of the $\gamma^*/Z$ boson produced at the Tevatron and decaying
into an $e^+e^-$ pair with invariant mass in the range
66~GeV~$<m_Z<$~116~GeV.
We compare the data with HERWIG before and after
matrix-element corrections; we also normalize the HERWIG distribution to the
experimental value of the cross section, 245.3~pb.
The result is that we obtain good agreement
with the experimental data only thanks to the application of the hard and soft
corrections, otherwise the predictions would have been badly wrong for
values $q_T>50$~GeV.
There is perhaps some evidence that HERWIG does not produce enough
smearing at low $q_T$, even with an intrinsic $p_t$ of 1~GeV, with
HERWIG peaking at about 2~GeV and the data peaking at about 3~GeV, but
the overall fit is nevertheless acceptable.

We have however found that better agreement can be obtained with an
intrinsic $p_t$ of 2~GeV.

\subsection{Jet distributions}

\begin{figure}[t]
\centerline{\resizebox{0.49\textwidth}{!}{\includegraphics{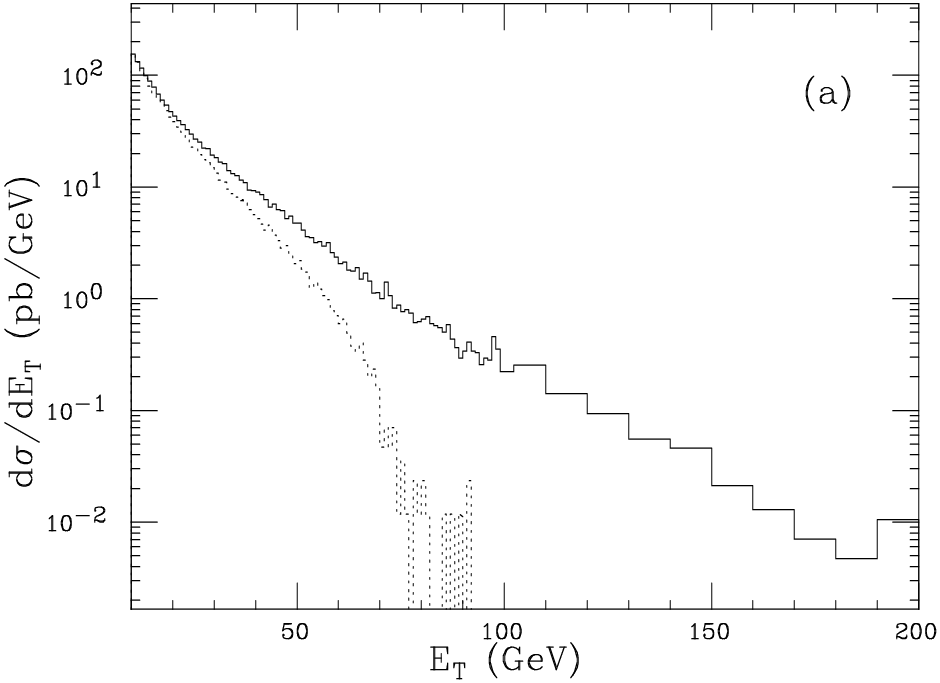}}%
\hfill%
\resizebox{0.49\textwidth}{!}{\includegraphics{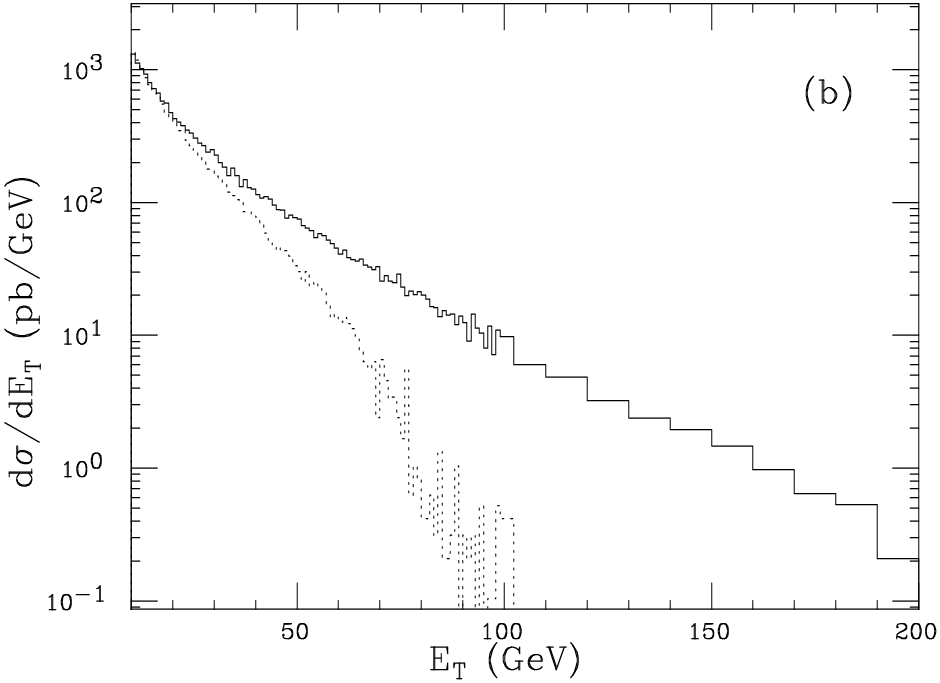}}}
\caption{The distribution in transverse momentum of the hardest jet in $W$
  production at the Tevatron (a) and the LHC (b) according to HERWIG
  without (dotted) and with (solid) matrix-element corrections.}
\label{fig:Wjet}
\end{figure}

We now look at the impact the matrix-element corrections have on
the jet activity at the Tevatron and at the LHC\@.
An interesting object to analyse
is the hardest jet in transverse energy (the so-called `first jet').
In Fig.~\ref{fig:Wjet} we plot the differential
spectrum for the transverse energy of the first jet for
$\sqrt{s}=1.8$~TeV and $\sqrt{s}=14$~TeV, using HERWIG 5.9
and 6.1 and running
the inclusive version of the $k_T$ algorithm [\ref{kt},\ref{soper}] for
a radius
$R=0.5$ at the Tevatron and $R=1$ for the LHC~\footnote{The correspondence
between the radius $R$ in the $k_T$ algorithm and $R_{\mathrm{cone}}$ in an
iterative cone algorithm is $R_{\mathrm{cone}}\approx0.75\times R$
[\ref{soper}].
The Tevatron experimentalists run an iterative cone algorithm with radius
$R_{\mathrm{cone}}=0.4$, so we choose $R=0.5$ for the radius parameter when
we consider jet events at $\sqrt{s}=1.8$~TeV.  For the LHC we stick to
the recommended value of $R=1$.}.
The result is that the improvement introduced does have a significant
effect
since the number of events in which the first jet has high $E_T$ is
significantly increased. The 5.9 and 6.1 distributions are similar for
small values of $E_T$, but for increasing $E_T$ the effect of the
corrections introduced in HERWIG gets more and more relevant and at very
high $E_T$ only events generated via the exact hard amplitude survive.
The impact is really enormous in the case of the LHC as can be seen
from Fig.~\ref{fig:Wjet}b.

\begin{figure}[t]
\centerline{\resizebox{0.49\textwidth}{!}{\includegraphics{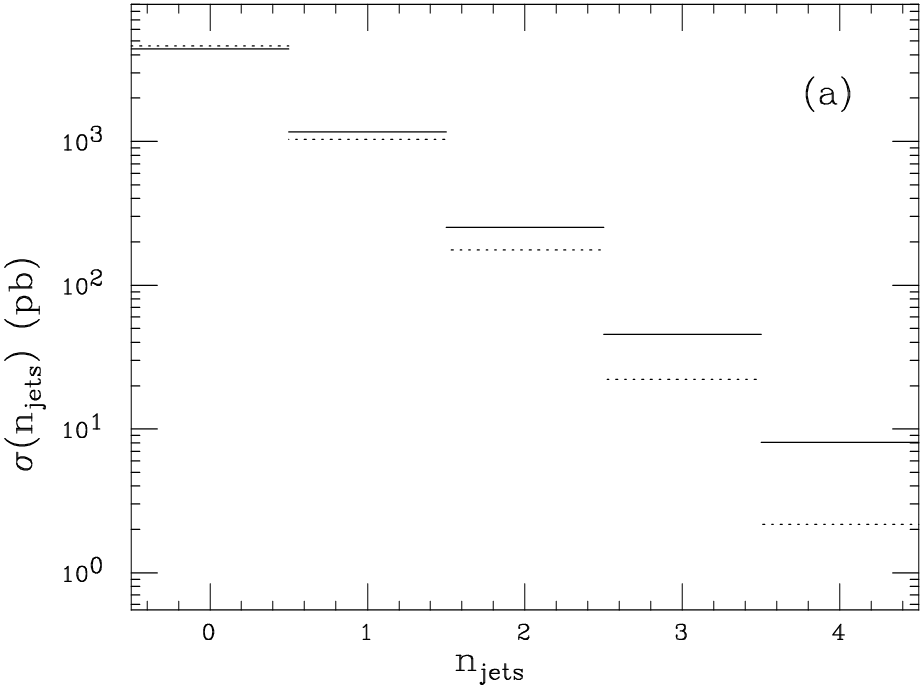}}%
\hfill%
\resizebox{0.49\textwidth}{!}{\includegraphics{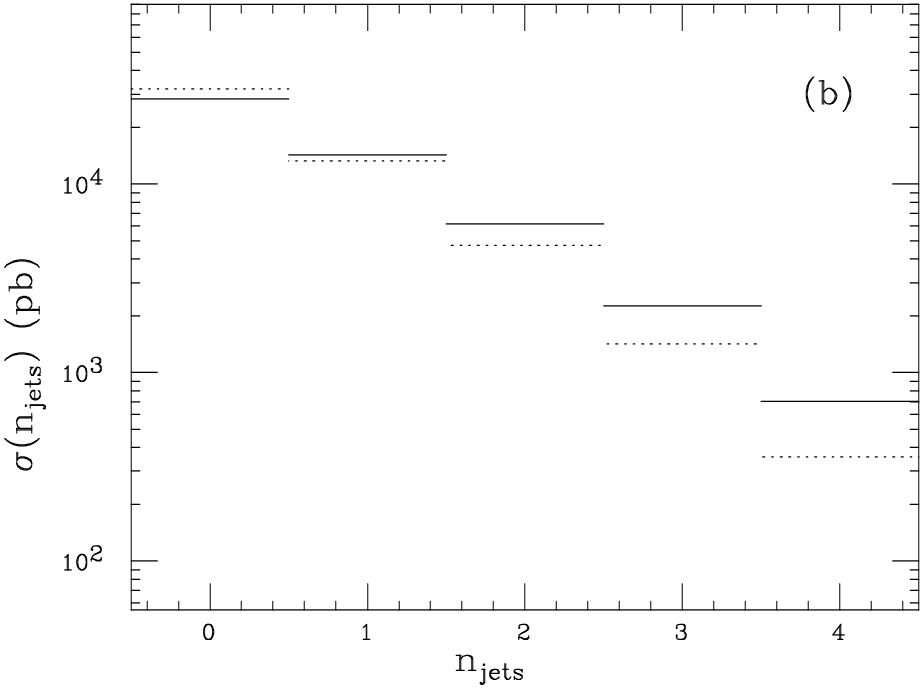}}}
\caption{The number of jets with \mbox{$E_T>10$~GeV} in $W$ events at the
  Tevatron (a) and the LHC (b) according to HERWIG without (dotted) and
  with (solid) matrix-element corrections.}
\label{fig:WNjets}
\end{figure}

In Fig.~\ref{fig:WNjets} we plot the inclusive number of jets
$n_{\mathrm{jets}}$ that pass a transverse energy cut $E_T>10$~GeV.
We see that implementing the matrix-element
corrections significantly shifts the distribution towards larger
$n_{\mathrm{jets}}$.
If we look at events with three or four jets having $E_T>10$~GeV, we see that
their number is increased considerably both at the Tevatron and at the LHC\@.
We have roughly an enhancement of a factor of 2 for three-jet events at both
the energies, while for events with four high transverse energy jets,
at the LHC we still get an enhancement of 2, while at the Tevatron
the difference is almost a factor of 4.

\subsection{Rapidity distributions}

\begin{figure}[t]
\centerline{\resizebox{0.49\textwidth}{!}{\includegraphics{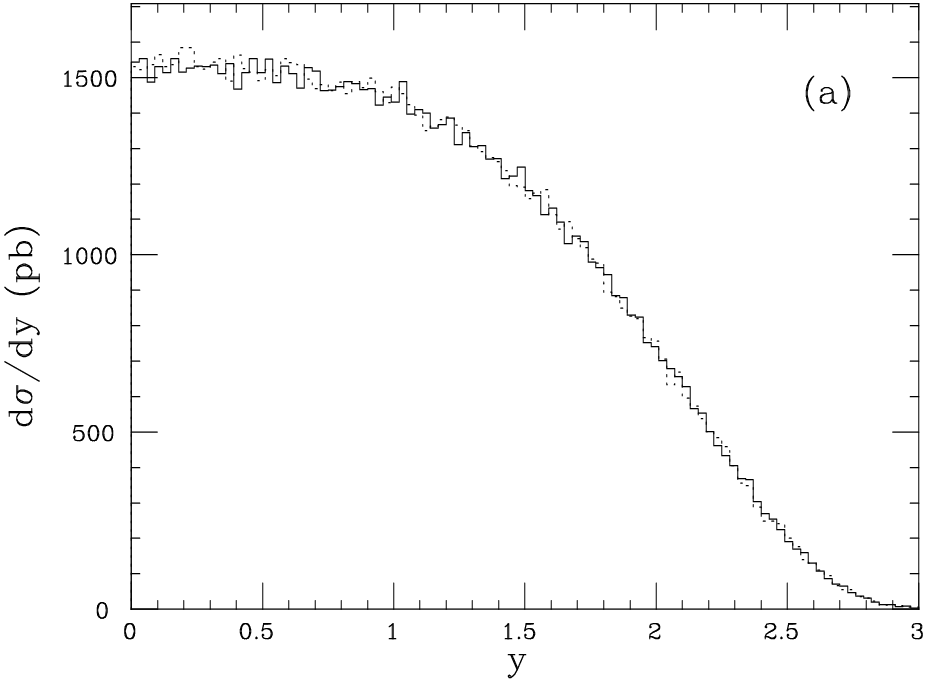}}%
\hfill%
\resizebox{0.49\textwidth}{!}{\includegraphics{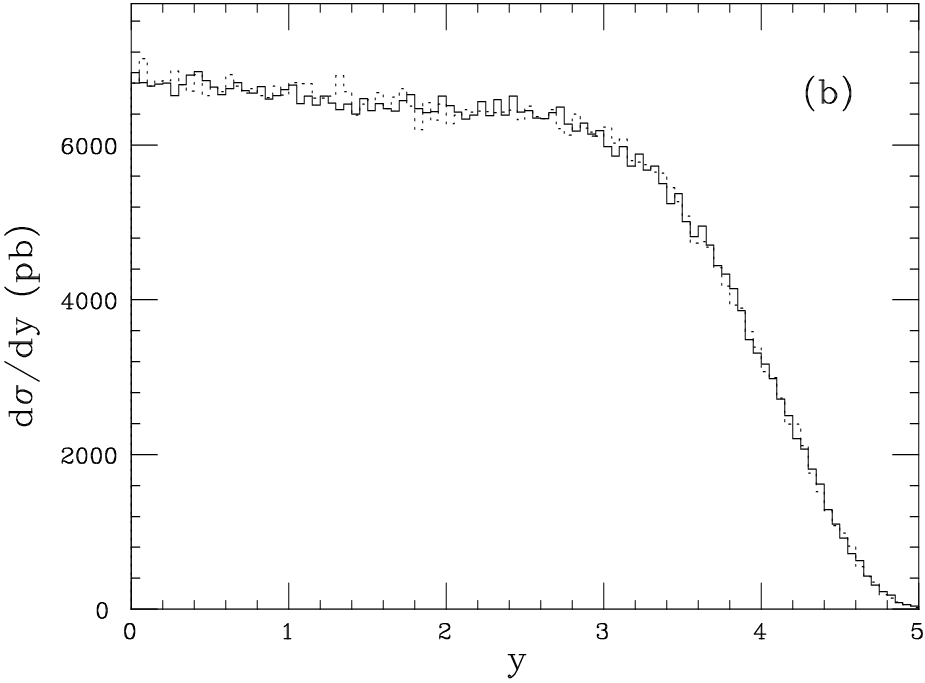}}}
\caption{The rapidity distribution of $W$ bosons at the Tevatron (a) and
  the LHC (b) according to HERWIG without (dotted) and with (solid)
  matrix-element corrections.}
\label{fig:Wrap}
\end{figure}

Fig.~\ref{fig:Wrap} shows the distribution of the rapidity $y$ of
the dilepton pair at $\sqrt{s}=1.8$~TeV and $\sqrt{s}=14$~TeV.
As we sum over $W^+$ and $W^-$ they are
symmetric in~$\pm y$. We see that the matrix-element corrections do not
significantly affect the rapidity distributions.

\begin{figure}[t]
\centerline{\resizebox{0.65\textwidth}{!}{\includegraphics{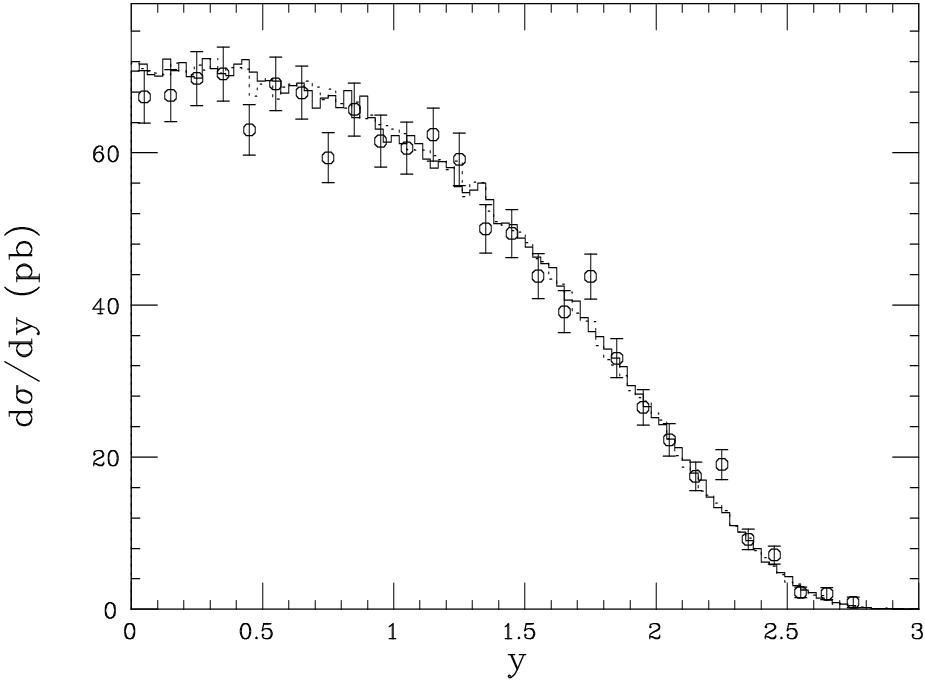}}}
\caption{The $Z$ rapidity distribution data from CDF, in comparison with
  HERWIG without (dotted) and with (solid) matrix-element corrections.}
\label{fig:ZrapCDF}
\end{figure}

Fig.~\ref{fig:ZrapCDF} shows the comparison between HERWIG and CDF [\ref{cdf}]
for the rapidity of the produced $e^+e^-$ pair; the agreement is again good
and the contribution of the matrix-element corrections is insignificant.

\section{Conclusions}

We have analysed Drell--Yan processes in hadron collisions in the
Monte Carlo parton shower approach.  This is accurate
in the soft/collinear approximation, but leaves empty regions in phase
space.
We implemented matrix-element corrections generating radiation according
to the
first-order amplitude in the dead zone and for every hardest so far emission
in the already populated region in the HERWIG parton shower.
We compared our results with the previous version HERWIG 5.9, with
experimental Tevatron data from the D\O\ and CDF collaborations and existing
resummed calculations of the spectrum of the transverse momentum $q_T$ of the
vector boson.
We found that the implemented corrections have a marked impact on the
phenomenological distributions for high values of $q_T$ and the
new version of HERWIG fits the D\O\ data for $W$ production well over the whole
$q_T$ spectrum after we correct the HERWIG results to detector level.
At large $q_T$ it is crucial to provide the Monte Carlo algorithm with
matrix-element corrections in order to succeed in obtaining such an agreement.

We also compared the HERWIG results after matrix-element corrections to some
existing calculations based on a resummation of the initial-state radiation
in the $q_T$-space and in the $b$-space.
We found that in the range of low $q_T$, where actually the effect of
matrix-element corrections is not so relevant and the resummed calculations are
quite reliable, the parton shower distribution
is in reasonable agreement with all of them, with discrepancies due to the
methods followed by these different approaches.
We also matched the resummed results to the exact ${\cal O}(\as)$ result,
in such a way to allow them to be trustworthy at all $q_T$ values and found that
the matching works well for a resummation performed in the $q_T$-space keeping
all the next-to-leading logarithms in the Sudakov exponent.
In this case, we also obtained good agreement with the HERWIG~6.1 $q_T$
distribution. The other approaches considered showed a discontinuity
at the point $q_T=m_W$ once we match them to the 
exact first-order perturbative result.

We also studied $W\;+$ jet events at the Tevatron and
at the LHC and found a significant effect of the new improvement of HERWIG
as a larger number of jets of large transverse energy passes the typical
experimental cuts.

We compared the new version of HERWIG with the experimental data
of the CDF collaboration on the transverse momentum and rapidity of $Z$
bosons. We found good agreement after implementing the corrections.
As a result, we feel confident that the simulation of vector boson production
is now reliable. Using the new version of HERWIG 6.1 to fit the experimental
data will therefore provide us with better tests of the Standard Model and of
QCD for the following Run~II at the Tevatron and, ultimately,
at the LHC\@.

For the sake of completeness we have however to say that our analysis has
been performed forcing the vector boson to decay into a lepton pair, as
most of the experimental studies do.
For hadronic channels (i.e.\ $W\to q\bar q'$), also the decay products are
allowed to emit gluons and to give rise to a parton shower that is still
described in the leading soft/collinear approximation by HERWIG\@.
The implementation of matrix-element corrections to hadronic
$W$ decays is straightforward as they are very similar to the
corrections to the process $Z\to q\bar q$ that are discussed in [\ref{sey2}]
and is in progress.
It is also worth remarking that the method applied in this paper to
implement matrix-element corrections to the initial-state radiation in
$W$ and $Z$ production can be extended to a wide range of processes that are
relevant for the phenomenology of hadron colliders.
Among these, the inclusion of matrix-element corrections to simulations of
heavy quark production and particularly of top production in $p\bar p$ or $pp$
interactions is expected to have a marked impact on the top mass
reconstruction and many observables that are relevant for heavy quark
phenomenology.
This work is also in progress.

\section*{Acknowledgements}

We acknowledge Stefano Frixione, Michelangelo Mangano, Stefano Moretti,
Willis Sakumoto and Bryan Webber for discussions of these and related topics.
We are indebted to Giovanni Ridolfi who provided us with the code to
obtain the plots in Fig.~\ref{fig:resum1}.
We are also grateful to the D\O\ Collaboration for making their detector
simulation, \texttt{CMS}, available to us, and especially to Cecilia
Gerber for the considerable effort it took to make it run outside the
usual D\O\ environment.

\section*{References}
\begin{enumerate}
\item\label{altarelli}
 G. Altarelli, R.K. Ellis and G. Martinelli, Nucl.\  Phys.\ B143 (1978) 521;
 Nucl.\ Phys.\ B146 (1978) 544 (erratum); Nucl.\ Phys.\ B157 (1979) 461.
\item\label{ddt}
 Yu.L. Dokshitzer, D.I. Dyakonov and S.I. Troyan, Phys.\ Rep.\ 58 (1980) 269.
\item\label{collins}
 J. Collins and D. Soper, Nucl.\ Phys.\ B193 (1981) 381; Erratum Nucl.\ Phys.\
 B213 (1983) 454; Nucl. Phys. B197 (1982) 446;\\
 J. Collins, D. Soper and G. Sterman, Nucl.\ Phys.\ B250 (1985) 199.
\item\label{ly}
  G.A. Ladinsky and C.P. Yuan, Phys.\ Rev.\ D50 (1994) 4239.
\item\label{davies}
 C.T.H. Davies, W.J. Stirling and B.R. Webber, Nucl.\ Phys.\ B256 (1985) 413.
\item\label{arnold}
 P.B. Arnold and R. Kauffman, Nucl.\ Phys.\ B349 (1991) 381.
\item\label{ellis1}
 R.K. Ellis, D.A. Ross and S. Veseli, Nucl.\ Phys.\ B503 (1997) 309.
\item\label{ellis2}
 R.K. Ellis and S. Veseli, Nucl.\ Phys.\ B511 (1998) 649.
\item\label{nason}
 S. Frixione, P. Nason and G. Ridolfi, Nucl.\ Phys.\ B542 (1999) 311.
\item\label{stirling}
 A. Kulesza and W.J. Stirling, DTP-99-02, hep-ph/9902234.
\item\label{pythia}
 T. Sj\"ostrand, Comp.\ Phys.\ Comm.\ 46 (1987) 367.
\item\label{herwig}
 G. Marchesini et al.,\ Comput.\ Phys.\ Commun.\ 67 (1992) 465.
\item\label{miu}
 G. Miu and T. Sj\"ostrand, Phys.\ Lett.\ B449 (1999) 313.
\item\label{mrenna}
 S. Mrenna, UCD-99-13, hep-ph/9902471.
\item\label{sey1}
 M.H. Seymour, Comput.\ Phys.\ Commun.\ 90 (1995) 95.
\item\label{sey2}
 M.H. Seymour, Z.\ Phys.\ C56 (1992) 161.
\item\label{sey3}
 M.H. Seymour, {\it Matrix Element Corrections to Parton Shower
    Simulation of Deep Inelastic Scattering}, contributed to 27th
  International Conference on High Energy Physics (ICHEP), Glasgow,
  1994, Lund preprint LU-TP-94-12, unpublished.
\item\label{corcella}
 G. Corcella and M.H. Seymour, Phys.\ Lett.\ B442 (1998) 417.
\item\label{sjostrand}
 T. Sj\"ostrand, Phys.\ Lett.\ B157 (1985) 231.
\item\label{marweb}
 G. Marchesini and B.R. Webber, Nucl.\ Phys.\ B310 (1988) 461.
\item\label{CMW}
 S. Catani, G. Marchesini and B.R. Webber, Nucl.\ Phys.\ B349 (1991) 635.
\item\label{sey4}
 M.H. Seymour, Nucl.\ Phys.\ B436 (1995) 443.
\item\label{mrs}
 A.D. Martin, R.G. Roberts and W.J. Stirling, Phys.\  Lett.\ B387 (1996) 419.
\item\label{d0}
 D\O\ Collaboration, B. Abbott et al., Phys.\ Rev.\ Lett.\ 80 (1998) 5498.
\item\label{CMS}
 D\O\ Collaboration, B. Abbott et al., Phys.\ Rev.\ Lett.\ 80 (1998) 3000;\\
 D\O\ Collaboration, B. Abbott et al., Phys.\ Rev.\ D58 (1998) 092003;\\
 I.~Adam, Ph.D. thesis, Columbia University, 1997, Nevis Report \#294,\\
 {\footnotesize\texttt{http://www-d0.fnal.gov/results/publications\_talks/thesis/adam/ian\_thesis\_all.ps}};\\
 E.~Flattum, Ph.D. thesis, Michigan State University, 1996,\\
 {\footnotesize\texttt{http://www-d0.fnal.gov/results/publications\_talks/thesis/flattum/eric\_thesis.ps}}.
\item\label{cdf}
 CDF Collaboration, T. Affolder et al., Fermilab-Pub-99/220-E;\\
 CDF Collaboration, A. Bodek et al., Fermilab-Conf-99/160-E.
\item\label{kt}
 S. Catani, Yu.L. Dokshitzer, M.H. Seymour and B.R. Webber,
 Nucl.\ Phys.\ B406 (1993) 187\\
 M.H. Seymour, Nucl.\ Phys.\ B421 (1994) 545.
\item\label{soper} S.D. Ellis and D.E. Soper, Phys.\ Rev.\ D48 (1993) 3160.
\end{enumerate}
\end{document}